\documentclass[aps,prl,twocolumn,reprint]{revtex4-1}
\usepackage{amsmath,amssymb,amsthm,bm,braket,ascmac,bbm,color}
\usepackage{graphicx}
\usepackage{comment,xcolor}

\theoremstyle{definition}

\newcommand{\mUF}{\mathcal{U}_F}

\newcommand{\mV}{\mathcal{V}}

\newcommand{\shp}{S_{\bm{h}}^+}
\newcommand{\shm}{S_{\bm{h}}^-}

\newcommand{\hL}{\hat{\mathcal{L}} }
\newcommand{\hU}{\hat{\mathcal{U}} }
\newcommand{\hD}{\hat{\mathcal{D}} }
\newcommand{\hH}{\hat{\mathcal{H}} }
\newcommand{\hA}{\hat{A} }
\newcommand{\hX}{\hat{X} }

\begin{document}
\title{
Time Crystals Protected by Floquet Dynamical Symmetry  in Hubbard Models}
\author{Koki Chinzei}
\affiliation{Institute for Solid State Physics, University of Tokyo, Kashiwa, Chiba 277-8581, Japan}

\author{Tatsuhiko N. Ikeda}
\affiliation{Institute for Solid State Physics, University of Tokyo, Kashiwa, Chiba 277-8581, Japan}
\date{\today}
\begin{abstract}
We investigate an unconventional symmetry in time-periodically driven systems,
the Floquet dynamical symmetry (FDS).
Unlike the usual symmetries, the FDS gives symmetry sectors that are equidistant in the Floquet spectrum and protects quantum coherence between them from dissipation and dephasing,
leading to two kinds of time crystals:
the discrete time crystal and discrete time quasicrystal
that have different periodicity in time.
We show that these time crystals appear in the Bose- and Fermi-Hubbard models under ac fields and their periodicity can be tuned only by adjusting the strength of the field.
These time crystals arise only from the FDS and thus appear in both dissipative and isolated systems and in the presence of disorder as long as the FDS is respected.
We discuss their experimental realizations in cold atom experiments
and generalization to the SU($N$)-symmetric Hubbard models.
\end{abstract}
\maketitle

{\it Introduction.}---
Symmetry is a key concept in physics,
presenting us with various information
such as conserved quantities, phase transitions and critical phenomena~\cite{Cardy1996}, and topological nature~\cite{Senthil2015}.
Even out of equilibrium, 
dynamics and nonequilibrium properties are governed by symmetries.
In particular, (time-)periodically driven (Floquet) systems~\cite{Holthaus2015,Bukov2015,Oka2019}
involve novel symmetries without equilibrium counterparts
due to the additional discrete time-translation symmetry,
and these symmetries give rise to, for example, 
the Floquet (discrete) time crystals~\cite{Else2016,VonKeyserlingk2016,Yao2017,Else2017,Zeng2017,Mizuta2018,Gong2018,Barberena2019,Lledo2019,Riera-Campeny2019,Zhu2019,Sacha2015,Russomanno2017,Ho2017,Giergiel2019,Choi2017,Zhang2017,Bordia2017,Rovny2018,Rovny2018a,Pal2018,Giergiel2018,Surace2019,Pizzi2019b,Pizzi2019,Giergiel2019b,Zhao2019}, Floquet symmetry protected topological phases~\cite{Oka2009,Kitagawa2010,Kitagawa2011,Jiang2011,Rudner2013,Potter2016,Kolodrubetz2018,McIver2020}, selection rules of high-harmonic generation in solids~\cite{Alon1998,Neufeld2019}, and so on~\cite{Berdanier2018}.

Unlike the usual symmetries leading to the conserved quantities,
there exist unconventional symmetries characterizing the nonequilibrium dynamics.
The dynamical symmetry~\cite{Buca2019,Medenjak2019,Tindall2020a,Medenjak2020,Munoz2019,Dogra2019,Buca2019b} in time-independent systems is one of them,
protecting some quantum coherence and leading to a time-crystalline state,
where the continuous time-translation symmetry $\mathbb{R}$ breaks down to the discrete one characterized by integers $\mathbb{Z}$~\cite{Wilczek2012,Li2012,Bruno2013,Bruno2013a,Watanabe2015,Buca2019,Medenjak2019,Tindall2020a,Nakatsugawa2017,Iemini2018,Kozin2019}.
The mechanism of this time crystal is different from the conventional Floquet time crystals, which occur in periodically driven systems.
The Floquet time crystals are characterized by the breaking of
the discrete time-translation symmetry $\mathbb{Z}$ down to
its subgroup such as $\mathbb{Z}/2$.

\begin{figure}[t]
\center
\includegraphics[width=1\columnwidth]{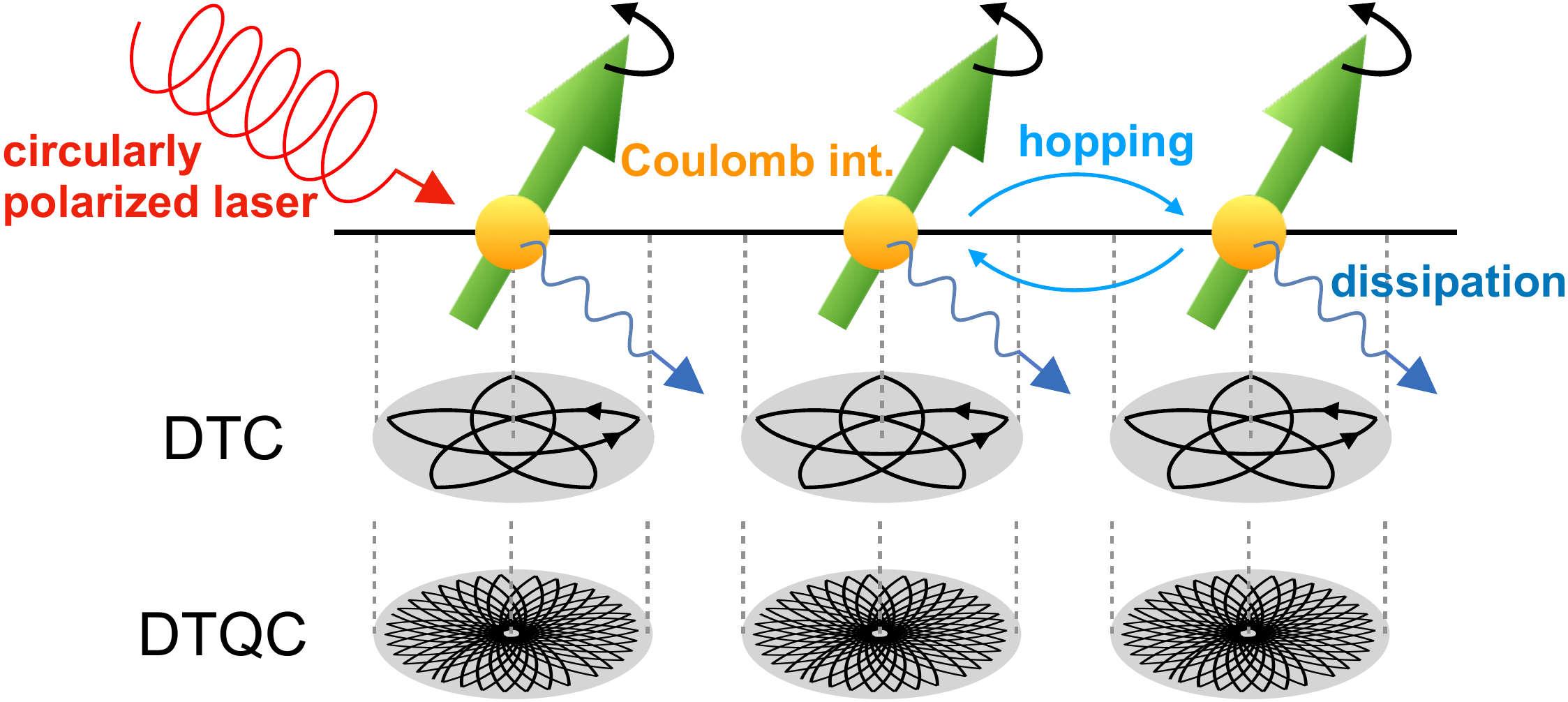}
\caption{
Schematic illustration of time crystals in Hubbard model.
The curves on the upper (lower) disks indicate the trajectories of the spin dynamics in the DTC (DTQC) phase.
}
\label{fig:illust}
\end{figure}

In this Letter, we investigate an unconventional symmetry in periodically driven systems, the Floquet dynamical symmetry (FDS),
showing that the FDS governs the long-time behavior of the system.
The FDS protects some quantum coherence in the Floquet spectrum from dissipation and dephasing
and leads to two kinds of time crystals:
the discrete time crystal (DTC) and discrete time quasicrystal (DTQC)~\cite{Pizzi2019,Giergiel2019b,Zhao2019}.
These time crystals both break the discrete time-translation symmetry,
but are different in that a perfect periodicity is retained or not (see Fig.~\ref{fig:illust}).
We show that these time crystals appear in various Hubbard models
under an ac field and their periodicity can be tuned only by adjusting the strength of the field.

{\it Floquet dynamical symmetry and time crystals.}---
We begin by considering periodically-driven dissipative systems
for a well-defined formulation and will discuss isolated systems later.
For this purpose,
we focus on
the Floquet-Lindblad master equation~\cite{Ho1986,Prosen2011,VorbergD13,Hartmann2017} ($\hbar = 1$ throughout this Letter):
\begin{align}
\frac{d\rho}{dt}
=\mathcal{L}_t(\rho) 
=
-i [H(t), \rho] + \gamma \sum_k \left( L_k \rho L_k^\dag - \frac{1}{2}\{L_k^\dag L_k,\rho\} \right), \label{eq:FLeq}
\end{align}
which describes trace-preserving nonunitary dynamics for the density matrix $\rho$.
Here, $H(t)=H(t+T)$ is a time-periodic Hamiltonian
describing the unitary part of the evolution,
and $L_k$'s are the Lindblad operators
representing the Markovian dissipation by an environment coupling to the system.

The solution of Eq.~\eqref{eq:FLeq} is formally given by
$\rho(t)=\mV(t,0)\rho(0)$, where $\mV(t,t')\equiv\mathcal{T}\exp(\int_{t'}^{t}\mathcal{L}_{s}ds)$ is the time evolution superoperator from $t'$ to $t$.
Due to the periodicity of $\mathcal{L}_t=\mathcal{L}_{t+T}$,
we can decompose $\mV(t,0)$ into a stroboscopic evolution and a micromotion:
$\mV(t,0)=\mV(\tilde{t},0)\mUF^\ell$,
where $t=\tilde{t}+\ell{T}$ \, ($0\leq\tilde{t}<T,\ell\in\mathbb{N}$),
and $\mUF=\mV(T,0)$ is the one-cycle  time evolution superoperator.
The long-time behavior ($\ell\rightarrow\infty$) is characterized by $\mUF$.
If $\mUF$ is diagonalizable $\mUF(\rho_j)=z_j\rho_j$ \, ($j=1,\cdots,D^2$, $D$ is the Hilbert-space dimension),
the formal solution of Eq.~\eqref{eq:FLeq} is given by $\rho(t) = \mV(\tilde{t},0)\sum_j a_{j}z_j^\ell\rho_j$, where $a_j$ is an expansion coefficients of $\rho(0)=\sum_j a_j\rho_j$.
Note that the trace-preserving nature guarantees $|z_j|\leq1$ 
and there exists at least one eigenvalue $z_j=1$, and thus we set $z_1=1$~\cite{Breuer2002}.

The unique nonequilibrium steady state (NESS) appears if all the other eigenvalues except $z_1=1$ satisfy
$|z_{j\neq 1}|<1$,
meaning that $\rho_{j\neq1}$ are all decaying modes with relaxation time $1/\ln|z_{j\neq1}|$~\cite{ikeda2020}.
In fact, any initial state asymptotically relax to a unique NESS,
$\rho(t)=\mV(\tilde{t},0)\sum_{j}a_{j}z_j^\ell{\rho_j}\xrightarrow{\ell\rightarrow\infty}\mV(\tilde{t},0)\rho_1$ ($a_1=1$ due to the trace preservation).
This NESS has time-period $T$ by definition of $t=\tilde{t}+\ell{T}$,
which means that the long-time behavior of Eq.~\eqref{eq:FLeq} typically has the discrete time-translation symmetry $\mathbb{Z}$.

The conventional discrete time crystals~\cite{Gong2018,Barberena2019,Lledo2019,Riera-Campeny2019,Zhu2019} appear if there exist several eigenstates with eigenvalues $e^{i\theta}\,(\theta=2\pi/n,\,n\in \mathbb{N})$.
For example, when $z_1=1$, $z_2=-1$, and $|z_{j\neq 1,2}| < 1$, 
any initial state asymptotically relax to the following NESS:
$\rho(t)=\mV(\tilde{t},0)\sum_{j}a_{j}z_j^\ell{\rho_j}\xrightarrow{\ell\rightarrow\infty}\mV(\tilde{t},0)[\rho_1+(-1)^\ell{a_2}\rho_2]$.
When $a_2 \neq 0$,
the NESS is a time-crystalline state with period $2T$,
which implies $\mathbb{Z}\rightarrow\mathbb{Z}/2$ symmetry breaking.
From now on, we call the eigenstate with eigenvalue 1 as a Floquet steady state, and that with eigenvalue $e^{i\theta}\,$($\neq 1$ with $\theta\in\mathbb{R}$) as a Floquet coherent state.

Now we introduce the Floquet dynamical symmetry (FDS),
which leads to unconventional time crystals.
First, we define the FDS 
for the unitary one-cycle time evolution operator $U_F=\mathcal{T}\exp[-i\int_0^T{ds}H(s)]$ 
and the dissipation $L_k$ as follows:
\begin{align}
\begin{aligned}
&\hspace{1cm}U_F A U_F^\dag = e^{-i\lambda T}A, \\
&[L_k,A(t)] = [L_k^\dag,A(t)]=0, \quad  \forall k,t \label{eq:FDS}
\end{aligned}
\end{align}
where $A(t)=U(t)AU^\dag(t)$ is an FDS operator at time $t$ ($ U(t)=\mathcal{T}e^{-i\int_0^t{ds}H(s)},A(0)=A,A(T)=e^{-i\lambda T}A$),
and $\lambda$ is a real number.
This definition is a natural extension of the strong dynamical symmetry in time-independent systems~\cite{Buca2019,Tindall2020a},
which is defined as  $[H,A]=\lambda{A}$ and $[L_k,A]=[L_k^\dag,A]=0$
(e.g., the Zeeman Hamiltonian $H=\lambda{S^z}$ and  the raising operator $A=S^+$ satisfy this relation,
$[H,A]=\lambda A$).
We note that the dissipation $L_k$ is not required for the FDS, and the time crystals can appear even in isolated systems, as we will show later.

The FDS protects some quantum coherence and prevents the system from relaxing to the unique NESS.
We remark that the FDS~\eqref{eq:FDS} implies $\mUF(A\rho)=e^{-i\lambda{T}}A\mUF(\rho)$ and  $\mUF(\rho{A}^\dag)=e^{i\lambda{T}}\mUF(\rho)A$ for any $\rho$ (see Supplemental Material S1).
Thus, given that $\rho_s$ is a Floquet steady state satisfying $\mathcal{U}_F(\rho_{s})=\rho_{s}$,
$\rho_{mn}=A^m\rho_{s}(A^\dag)^n$ are the Floquet steady ($m=n$) and coherent ($m\neq{n}$) states,
\begin{align}
\mUF(\rho_{mn}) = e^{i(n-m)\lambda T} \rho_{mn}. \label{eq:UFrhomn}
\end{align}
If there are no other Floquet steady and coherent states except $\rho_{mn}$,
we obtain the long-time behavior from them,
\begin{align}
\rho(t) 
\xrightarrow{\ell \rightarrow \infty} \sum_{mn} \mV(\tilde{t},0) [ c_{mn} e^{i (n-m)\lambda \ell T} \rho_{mn}]
\equiv \rho_{\infty}(t),\label{eq:long-time}
\end{align}
where $c_{mn}$ is expansion coefficients of $\rho_{mn}$.
There exist two typical energy (or time) scales in $\rho_{\infty}(t)$.
One is the Floquet frequency $\omega=2\pi/T$ stemming from the periodicity of $\mV(\tilde{t},0)$,
and the other is $\lambda$ characterized by the FDS.

Depending on whether the ratio $\lambda/\omega$
is a rational number or not,
the long-time behavior~\eqref{eq:long-time}
represents the DTC or DTQC, respectively.
If $\lambda/\omega\in\mathbb{Q}$,
i.e., $\lambda/\omega=q/p$ for some coprime integers $p$ $(\ge2)$ and $q$,
the DTC emerges:
$\rho_{\infty}(t+T)\neq\rho_{\infty}(t)$, but
$\rho_{\infty}(t+pT)=\rho_{\infty}(t)$ holds true.
This means the discrete time-translation symmetry breaking, $\mathbb{Z}\rightarrow\mathbb{Z}/p$.
On the other hand, if $\lambda/\omega\notin\mathbb{Q}$,
the discrete time-translation symmetry $\mathbb{Z}$ is broken,
but
there is no integer $p$ such that $\rho_{\infty}(t+pT)=\rho_{\infty}(t)$.
Nevertheless, there exist an infinite number of times
$s$ such that $\rho_{\infty}(t+s)$
is arbitrarily close to $\rho_{\infty}(t)$.
Thus, the dynamics is quasiperiodic, and we call it the DTQC.
Note that these time crystals are protected by the FDS
and robust against the perturbations respecting the symmetry.

{\it Tunable time crystals in dissipative Hubbard models.}---
Here we present simple models exhibiting the time crystals protected by the FDS: 
spin-$S$ Bose- or Fermi-Hubbard models under a circularly polarized ac field
\footnote{
We ignore the coupling between the electric field and electric charges for simplicity.
If included, this coupling does not change the results qualitatively
since it does not affect the spin dynamics
}
in $d$ dimensions.
The Hamiltonian $H(t)=H_0 + V(t)$ is 
\begin{align}
\begin{aligned}
&H_0
=-J\sum_{\braket{i,j},\sigma} (c^\dag_{i,\sigma} c_{j,\sigma} + \text{h.c.})  
 + \frac{U}{2} \sum_j n_{j}^2  
+\frac{K}{2}\sum_{\braket{i,j}} n_{i} n_{j},   \\
&V(t)
=B \left( S^x \cos\omega t + S^y \sin\omega t \right), \label{eq:hubbard}
\end{aligned}
\end{align}
where  $c_{j,\sigma}(c_{j,\sigma}^{\dag})$ is the annihilation (creation) operator for the boson or fermion with spin $\sigma \in \{ -S, \cdots, S \}$ on the site $j$,
and $n_{j}=\sum_{\sigma} c_{j,\sigma}^{\dag}c_{j,\sigma}$ is the particle number operator.
The operators for the spin at site $j$ and for the total spin are denoted by $S_j^\mu$ and $S^\mu=\sum_j S_j^\mu \, (\mu = x,y,z)$.
The nearest-neighbor interaction is added to break the integrability of the model in $d=1$.

We consider the dissipation described by local dephasing Lindblad operators acting on each site, $L_j=n_j$,
which suppresses the particle number fluctuation.
We will discuss later how to realize them experimentally.

\begin{figure*}[t]
\center
\includegraphics[width=\linewidth]{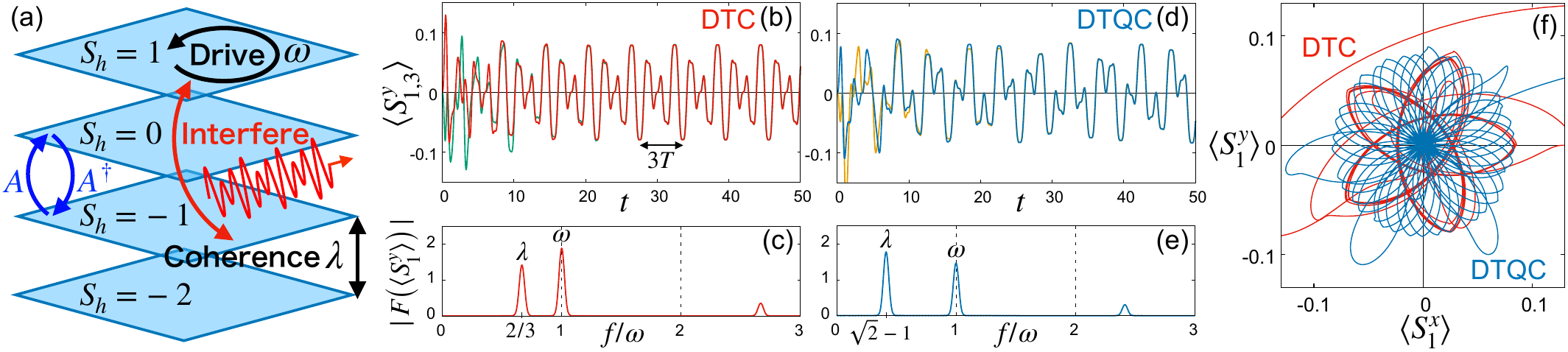}
\caption{
(a) Mechanism of time crystals by Floquet dynamical symmetry.
Each sheet represents the different spin sectors.
The periodic drive and the quantum coherence interfere with each other.
(b) DTC time evolutions of $\braket{S^y_j(t)}$ (red: $j=1$, green: $j=3$)
and (c) the Fourier component ($j=1$).
(d) DTQC time evolutions of $\braket{S^y_j(t)}$ (blue: $j=1$, yellow: $j=3$)
and (e) the Fourier component ($j=1$).
We have performed  the Fourier transformation for $t \in [20,100]$ by convoluting the window function $w(t)=\exp[-(t-60)^2/20^2]$.
(f) Trajectories of time evolution in $(S_1^x,S_1^y)$-plane. The red (blue) curve denotes the DTC (DTQC) trajectory.
The parameters are $L=6,J=U=K=1,\gamma=0.1,\omega=\pi$, and $B=4\pi/3$ ($B=\pi$) for the DTC (DTQC). 
}
\label{fig:main}
\end{figure*}

Our model has the FDS in the one cycle of the ac field.
To show this, we consider a unitary transformation to the rotating frame~\cite{Slichter1996,Kohler2005,Takayoshi2014,Zhao2019}, 
$\rho_\text{RF}(t)=R(t)\rho(t)R(t)^\dag$ ($R(t)=e^{i\omega{t}S^z}$).
Then, the Hamiltonian is transformed to the Hubbard model in an effective static magnetic field,   $H_\text{RF} = R(t) [H(t) - i \partial_t] R(t)^\dag = H_0 + \bm{h} \cdot \bm{S}$, where $\bm{h}=(B,0,\omega)$,
while the Lindblad operators are invariant.
Therefore, in the rotating frame,  the system has the strong dynamical symmetry~\cite{Buca2019,Tindall2020a},
or the FDS $U_F^\text{RF} A  (U_F^\text{RF})^\dag = e^{-i\lambda T} A $ and $[L_j,A]=[L_j^\dag,A]=0$,
where $U_F^\text{RF}$ is  the unitary one-cycle  time evolution operator in the rotating frame,
$A=S_{\bm{h}}^+$ is the total spin raising operator along the effective magnetic field $\bm{h}$, and 
\begin{align}
\lambda = |\bm{h}| = \sqrt{\omega^2 + B^2}  \quad (\text{mod} \; \omega).
\end{align}
Going back to the original frame, we have $U_F = R^\dag(T) U_F^\text{RF} R(0) = U_F^\text{RF}$  up to the phase factor
and $A(t)=R^\dag(t) AR(t)=S^+_{\bm{h}(t)}$, 
where $S^+_{\bm{h}(t)}$ is the raising operator along $\bm{h}(t)=(B\cos \omega t, B\sin \omega t,\omega)$,
obtaining the FDS
\begin{align}
\begin{aligned}
&\hspace{1cm}U_F \shp U_F^\dag = e^{-i\lambda T} \shp, \\
&[L_j, S^+_{\bm{h}(t)}]=[L_j^\dag, S^+_{\bm{h}(t)}]=0, \quad \forall j,t.
\end{aligned}
\label{eq:FDShubbard}
\end{align}
We remark that the FDS also exists under elliptically-polarized fields for spin-$1/2$ Fermi-Hubbard model,
for which we cannot find a rotating frame giving a static Hamiltonian~(see Supplemental Material S2).
Thus, the FDS is a more general concept not restricted to the rotating-frame argument.

The FDS~\eqref{eq:FDShubbard}
gives the Floquet steady and coherent states $\rho_{mn}=(\shp)^m \rho_{s} (\shm)^n$ with eigenvalue $e^{i(n-m)\lambda T}$ in Eq.~\eqref{eq:UFrhomn}.
This implies that the FDS protects the quantum coherence between the different spin sectors labelled by $S_{\bm{h}}=\bm{h} \cdot \bm{S}/|\bm{h}|$, which are Zeeman-split by $\lambda$ in the Floquet spectrum (see Fig.~\ref{fig:main} (a)). 
Meanwhile, the quantum coherence within each sector
are eliminated by dissipation (and quantum thermalization discussed below),
and the Floquet steady state oscillating with frequency $\omega$ appears within each sector.
These Floquet steady states are superposed and the two scales, $\lambda$ and $\omega$, interfere with each other,
giving rise to  the DTC and DTQC.

We emphasize the tunability of our time crystals.
As shown before, the periodicity of these time crystals depend on the ratio $\lambda/\omega=\sqrt{1+(B/\omega)^2}$,
which can be tuned only by varying the field strength $B$ with $\omega$ fixed.
If $\lambda/\omega = q/p$ ($\leftrightarrow$ $B=\omega \sqrt{(q/p)^2-1}$) with coprime integers $p$ and $q$, the DTC appears with period $pT$,
whereas, if $\lambda/\omega \notin \mathbb{Q}$, the DTQC appears.

{\it Synchronized DTC and DTQC in the Hubbard model.}---
Let us numerically demonstrate the DTC and DTQC 
by taking the spin-1/2 Fermi-Hubbard model in one dimension with $L$ sites.
We assume the periodic boundary condition and solve Eq.~\eqref{eq:FLeq} by the fourth-order Runge--Kutta method.
Throughout this Letter, the initial state is a $1/4$-filled state where the $j$-th site ($j=1,2, \cdots, L/2$) is occupied by one fermion,
and every third fermions are polarized along $-x$ and all others are polarized along $x$ (we assume $L$ is a multiple of 6).

Figure~\ref{fig:main} (b) shows the time evolution of $\braket{S_j^y(t)} (j=1,3)$ in the DTC phase for $\omega = \pi$ and $\lambda = 2\pi/3$.
After the initial relaxation dynamics, $\braket{S_j^y(t)}$ oscillates with period $3T = 6$, which is the least common multiple of $T=2\pi/\omega = 2$ and $2\pi/\lambda = 3$. 
This means the discrete time-translation symmetry breaking $\mathbb{Z} \rightarrow \mathbb{Z}/3$.
The rationality of two peaks in the Fourier space (Fig.~\ref{fig:main} (c)) at $f = \omega$ and $\lambda$ also
shows the commensurability.
Moreover, the dynamics of $\braket{S_1^y(t)}$ and $\braket{S_3^y(t)}$ are synchronized after relaxation,
which implies all the Floquet steady and coherent states are translationally symmetric~\cite{Buca2019,Tindall2020a}.

On the other hand, Fig.~\ref{fig:main} (d) shows the time evolution in the DTQC phase for $\omega = \pi$ and $\lambda = (\sqrt{2}-1)\pi$,
where the synchronized spins oscillate aperiodically.
The irrationality of two peaks,
$f = \omega$ and $\lambda$,
in the Fourier space (Fig.~\ref{fig:main} (e)) shows the incommensurability of the time crystal
and no perfect periodicity.

The trajectories of the DTC and DTQC dynamics in the $(S^x_1, S^y_1)$-plane are shown in Fig.~\ref{fig:main} (f).
The trajectory of the DTC dynamics (red) behaves as the limit cycle,
and gradually converges to the star-shaped closed curve.
On the other hand, the trajectory of the DTQC dynamics (blue) never converges to a closed curve,
and keep rotating aperiodically on the plane.
This aperiodic dynamics highlights the difference from the DTC.

{\it Quantum-thermalization-induced time crystals without dissipation.}---
Until now, we have considered the dissipative systems to clarify the role of the FDS and the mechanism of the time crystals.
However, the time crystals do not necessarily require the dissipation.
According to the recent studies~\cite{DAlessio2016,Eisert2015a,Mori_2018},
an isolated quantum system without dissipation exhibits thermalization
(or, more precisely, equilibration) due to dephasing
between many-body energy eigenstates
when the system size is large enough~\cite{Tasaki1998,Reimann2008,Short2011}.
The quantum thermalization (equilibration) effectively plays the role of dissipation and eliminates quantum coherence except those protected by the FDS, bringing about the time crystals
\footnote{
See Supplemental Material S3 for time crystals in integrable systems, where the equilibration plays the important role instead of the thermalization.
}.

\begin{figure}[t]
\center
\includegraphics[width=\columnwidth]{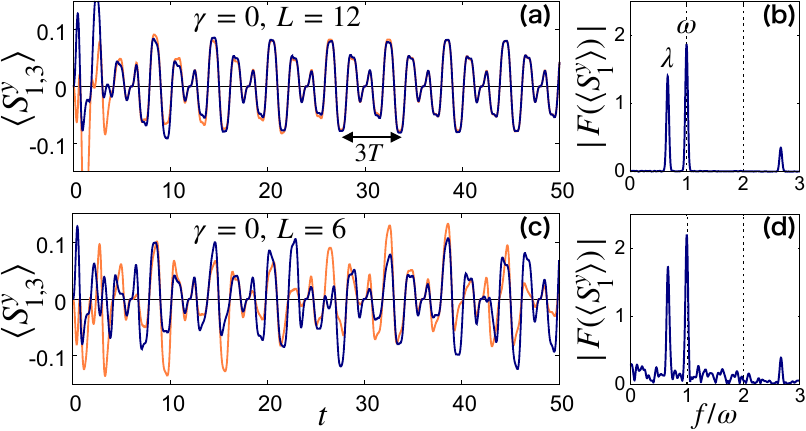}
\caption{
Time profile $\braket{S^y_{j}(t)}$ (left)
and its Fourier spectrum (right)
calculated in the absence of dissipation ($\gamma=0$).
The upper (lower) panels correspond to $L=12$ ($L=6$),
and the navy (orange) curve to site $j=1$ ($j=3$).
The other parameters are $J=U=K=1, \omega=\pi$, and $B=4\pi/3$ (DTC phase).
}
\label{fig:largesmall}
\end{figure}

We demonstrate the time crystal protected by the FDS without dissipation (i.e., $\gamma=0$)
in Fig.~\ref{fig:largesmall}.
Both in the time profile and Fourier spectrum,
we observe a clear time-crystalline behavior at $L=12$ (upper panels)
whereas a noisy one at $L=6$ (lower panels).
This noise derives from the imperfect dephasing
as a finite-size effect
and typically decreases exponentially in the system size.

The time crystals in isolated systems
are interpreted by the maximum entropy principle
under multiple conserved quantities~\cite{Jaynes1957}
that is also known as the generalized Gibbs ensemble (GGE)~\cite{Rigol2007,Medenjak2019}~\footnote{
The terminology GGE is sometimes limited to the case when
there exists an extensive number of conserved quantities, unlike our model.
}.
The key to this interpretation is the unconventional stroboscopically-conserved quantities
that are derived from the FDS.
To show this intuitively,
we focus on the DTC case with $\lambda/\omega=q/p$
$(p\ge2)$.
One can easily show that 
the FDS $U_F A U_F^\dag = e^{-i\lambda T}A$ leads to such quantities after $p$ steps:
 \begin{align}
[U_F^p, A] = [U_F^p, A^\dag] = 0.
\end{align}
These stroboscopically-conserved quantities, $A$ and $A^\dag$,
constrain the dynamics and
lead, after a long time, to the GGE at $t=npT$ ($n\in\mathbb{N}$):
$\rho_\text{TC}(t=npT) 
= \exp \left[ -\sum_i \beta_i Q_i - \mu A - \mu^\ast A^\dag \right]/Z$.
Here $Z$ is the partition function,
$\mu$ is the chemical potential for $A$,
and $Q_i$ and $\beta_i$ are the conventional local conserved quantities of $U_F$
and their generalized inverse temperatures.
Acting the time evolution operator $U(t_0)$ from $t=npT$ to $t=t_0+npT$ ($t_0\in[0,pT)$),
we obtain the time-dependent GGE $\rho_\text{TC}(t_0+npT) = U(t_0) \rho_\text{TC}(npT)U^\dag(t_0)$:
 \begin{align}
\rho_\text{TC}(t) 
= \frac{\exp \left[ -\sum_i \beta_i Q_i(t) - \mu(t) \tilde{A}(t) - \mu(t)^\ast \tilde{A}^\dag(t) \right]}{Z}, \label{eq:tFGGE} 
\end{align}
where  $Q_i(t)=U(t) Q_i U^\dag(t)$, $\tilde{A}(t)=e^{i\lambda t}U(t) A U^\dag(t)$, and $\mu(t) = \mu e^{-i\lambda t}$.
Remarkably, under a certain assumption,
this result~\eqref{eq:tFGGE} also holds in the DTQC case where there exist no stroboscopically-conserved quantities (see Supplemental Material S4).

The time-dependent GGE~\eqref{eq:tFGGE} is a two-color generalization of the previous ones~\cite{Lazarides2014,Medenjak2019}.
Whereas $Q_i(t)$ and $\tilde{A}(t)$ have the period $T=2\pi/\omega$ of the external field,
$\mu(t)$ has a different one $T_\lambda=2\pi/\lambda$ of the FDS.
These two periods, depending on their ratio, give the DTC and DTQC in isolated systems. 
The synchronization implies the translation symmetry of $Q_i(t)$ and $\tilde{A}(t)$.

In deriving Eq.~\eqref{eq:tFGGE},
we have implicitly assumed that there are no other FDS than $A$.
This assumption is known to break down in free and many-body-localized systems that have an extensive number of local dynamical symmetries.
In these systems, 
the dense and incommensurate frequencies associated with the multiple FDSs can destroy time crystals~\cite{Essler2016,Khemani2019,Bokker2020}.
It is an open question to elucidate the roles of these multiple FDSs.

Finally, we remark the robustness of our time crystals against disorder.
As shown above, the time-crystalline nature relies only on the FDS, and the disorder never disturbs the time crystals as long as it respects the FDS (and there exist no other FDS).
For example,
random onsite potentials~\cite{Prelovsek2016},
which are SU($2$)-symmetric unlike random onsite magnetic fields,
do not destroy 
the time crystals both in the dissipative and isolated systems
(see Supplemental material S5).

{\it Discussions and conclusions.}---
In this Letter, we have introduced the FDS and proposed a new class of time crystals protected by the FDS.
Although we have focused on the circularly polarized ac field,
the time crystals also appear in the Hubbard model with static and linearly-polarized ac fields both along the $z$-direction.
This is because a trivial dynamical symmetry holds at any time, $[H(t),S^+] \propto S^+$.
This setup is known as the electron spin resonance (ESR) in the condensed matter physics~\cite{Weil1984}.

Our model~\eqref{eq:hubbard} can be realized in ultracold atoms on an optical lattice~\cite{Jaksch2005,Esslinger2010}. 
The hyperfine states of the atom behave as a pseudospin, and the coupling between the states, or the Zeeman effect on the pseudospin, is manipulated by radio waves~\cite{Esslinger2010}.
State-of-the-art laser technology enables us to control the strength and frequency of the coupling,
realizing the highly tunable time crystals.
Also, the particle number dephasing $L_j=n_j$ is achieved by immersing the optical lattice into a Bose-Einstein condensate~\cite{Klein2007,Mayer2019}.

Finally, we note that our model~\eqref{eq:hubbard} actually has the SU($N$) symmetry
with $N=2S+1$~\cite{Honerkamp2004,Cazalilla2009a,Gorshkov2010a}.
Such an SU($N$)-symmetric Hubbard model has been realized in ultracold atoms~\cite{Taie2010,Zhang2014d,Scazza2014},
and has attracted much attention recently.
This model could accommodate more exotic time crystals with multiple input frequencies and the extended FDS due to the enlarged algebraic structure.
Generalizing our arguments to the SU($N$) systems
is an interesting open question, 
where time crystals should meet large Lie algebras together with cold atom experiments.

{\it Acknowledgements.}---
We thank K. Fukai and H. Tsunetsugu for fruitful discussions. 
This work was supported by JSPS KAKENHI Grant No.~JP18K13495.
K. C. acknowledges the financial support provided
by the Advanced Leading Graduate Course for Photon Science at the University of Tokyo.

{\it Note added.}---
In preparing this manuscript,
we have become aware of an independent work~\cite{Medenjak2020},
where the FDS and its consequences are briefly discussed.

%\bibliography{../../../../../bibtex/dynamical_symmetry}%#### for chinzei


%#####################################################################
%#####################################################################
%#####################################################################
%式番号をサプリ用に変更
\setcounter{equation}{0}
\makeatletter
\def\tagform@#1{\maketag@@@{(S#1)}}
\makeatother
%####################

%図番号をサプリ用に変更
\setcounter{figure}{0}
\renewcommand{\thefigure}{S\arabic{figure}}

%###################
%引用番号をサプリ用に変更
%\makeatletter
%\renewcommand{\@cite}[2]{[S#1]}
%\renewcommand{\@biblabel}[1]{[S#1]}
%\makeatother
%###################

\newpage
\vspace{0.5cm}
\begin{center} {\large {\bf Supplemental Materials}} \end{center}
\vspace{-0.5cm}

\section{S1. Derivation of Floquet steady and coherent states}

Here we briefly prove that the Floquet dynamical symmetry (FDS) leads to 
\begin{align}
\begin{aligned}
&\mUF(A\rho) = e^{-i\lambda T} A \mUF(\rho), \\
&\mUF(\rho A^\dag) = e^{i\lambda T} \mUF(\rho) A, 
\end{aligned} \label{sp:hosii}
\end{align}
for any $\rho$, and thus the time crystals (see Eqs.~(3) and (4) in the main text).
To this end, let us introduce superoperators $\hat{O}_L$ and $\hat{O}_R$ ($O$ is a usual operator),
which act on the density matrix from the left and right sides such that
$\hat{O}_L \rho = O\rho$ and $\hat{O}_R \rho = \rho O^\dag$.

Using FDS superoperators $\hA_L$ and $\hA_R$, Eq.~\eqref{sp:hosii} reads
\begin{align}
\begin{aligned}
&\hU_F \hA_L = e^{-i\lambda T}\hA_L \hU_F,  \\
&\hU_F \hA_R = e^{i\lambda T}\hA_R \hU_F, 
\end{aligned}\label{sp:kankei}
\end{align}
where $\hU_F = \mathcal{T} \exp[\int_0^T \hL_s ds]$ is the one-cycle time evolution superoperator,
and the Liouvillian $\hL_t$ is a sum of the unitary part $\hH_t$ and the dissipative one $\hD$, $\hL_t=\hH_t+\hD$.
Note that the FDS $[L_k, A(t)]=[L_k^\dag,A(t)]=0$ leads to commutation relations $[\hD,\hA_L(t)]=[\hD,\hA_R(t)]=0$.

To prove the upper one in Eq.~\eqref{sp:kankei} (the lower one is also proven by a similar argument),
we consider a superoperator $\hX(t) = \hU(t) \hA_L \hU(t)^{-1}$, 
where $\hU(t) = \mathcal{T} \exp[\int_0^t \hL_s ds]$ is the time evolution superoperator from 0 to $t$, 
and $\hU(t)^{-1} = \tilde{\mathcal{T}} \exp[-\int_0^t \hL_s ds]$ is its inverse ($\tilde{\mathcal{T}}$ denotes the anti-time-ordering operator).
By using the relations $\partial_t \hU(t) = \hL_t \hU(t)$ and $\partial_t \hU(t)^{-1} = -\hU(t)^{-1} \hL_t$,
we have a  time evolution equation of $\hX(t)$: 
\begin{align}
\partial_t \hX(t) = [\hL_t, \hX(t)].
\end{align}
One can easily show that the solution of this equation is $\hX(t) = \hA_L(t)$
because of $\partial_t \hA_L(t) = [\hH_t,\hA_L(t)]$ and $[\hD,\hA_L(t)]=0$ that is derived by the FDS.
Therefore, 
we obtain Eq.~\eqref{sp:kankei} by $\hU_F \hA_L \hU_F^{-1}=\hX(T) = \hA _L(T) = e^{-i\lambda T}\hA_L$,
and thus Eq.~\eqref{sp:hosii},
where we have used the FDS, $A(T)=U_F A U_F^\dag = e^{-i\lambda T} A$.

\begin{figure*}[t]
\center
\includegraphics[width=\linewidth]{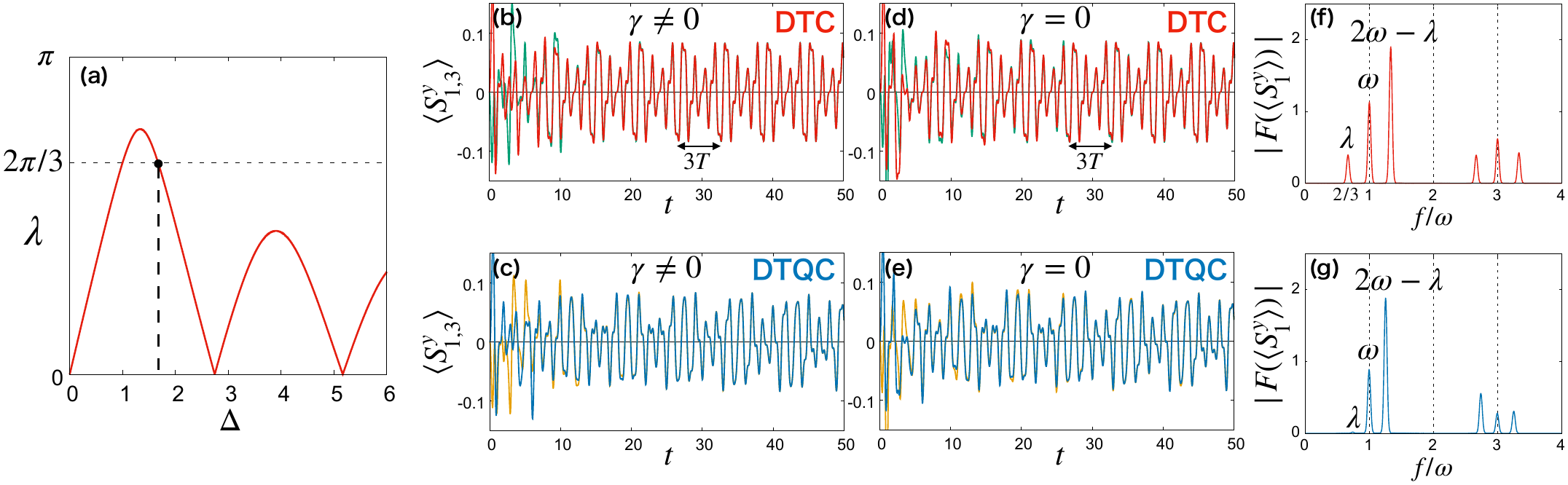}
\caption{
(a) Strength of effective magnetic field for $\omega=\pi$ and $B=4\pi/3$ with varied $\Delta$.
(b,d) DTC time evolutions of $\braket{S^y_j(t)}$ (red: $j=1$, green: $j=3$) at $\Delta \simeq 1.673$ ($\lambda=2\pi/3$) in (b) the dissipative and (d) isolated systems.
(c,e) DTQC time evolutions of $\braket{S^y_j(t)}$ (blue: $j=1$, yellow: $j=3$) at $\Delta = 1.5$ in (c) the dissipative and (e) isolated systems.
The parameters are $J=U=K=1, \omega=\pi, B = 4\pi/3$, and $\gamma=0.1$ for (b,c) ($\gamma=0$ for (d,e)).
The system sizes are $L=6$ for (b,c) and $L=12$ for (d,e).
(f,g) Fourier components of $\braket{S^y_1(t)}$ in (b) and (c) respectively.
We have performed  the Fourier transformation for $t \in [20,100]$ by convoluting the window function $w(t) = \exp[-(t-60)^2/20^2]$.
}
\label{fig:elli}
\end{figure*}

\section{S2. Hubbard model under elliptically polarized ac field}

Here, we show that the spin-1/2 Fermi-Hubbard model under an elliptically polarized ac field also possesses the FDS and exhibits the time-crystalline behaviors.
The Hamiltonian $H_\text{el}(t) = H_0 + V_\text{el}(t)$ is given by 
\begin{align}
&
\begin{aligned}
H_0 &= -J\sum_{\braket{i,j}\sigma} (c^\dag_{i\sigma} c_{j\sigma} + \text{h.c.}) \\ 
 &\hspace{0.5cm} + (U/2) \sum_j n_{j}^2 + (K/2)\sum_{\braket{i,j}} n_{i} n_{j}, 
\end{aligned} \\
&V_\text{el}(t) = \sum_j V_j(t) = B \left( \Delta S^x \cos\omega t + S^y \sin\omega t \right),
\end{align}
where $V_j = B \left( \Delta S^x_j \cos\omega t + S^y_j \sin\omega t \right)$
denotes the Zeeman coupling to the elliptically polarized field
at the $j$-th site.
The parameter $\Delta$ characterizes the ellipticity, and $\Delta=1$ corresponds to the circular polarization.

To show the FDS,
we consider the one-cycle time evolution operator under $H_\text{el}(t)$, $U_F^\text{el} = \mathcal{T} \exp[-i \int_0^T dt H_\text{el}(t)]$.
By using the commutation relations $[H_0, V_\text{el}(t)]=0$ and $[V_i(t),V_j(t')]=0 \,\, (i \neq j)$,
we have $U_F^\text{el} = e^{-iH_0T} \bigotimes_j U_j$, where $U_j = \mathcal{T} \exp[-i \int_0^T dt V_j(t)]$ is a unitary operator acting on the $j$-th site,
where the Hilbert space is 4-dimensional.
We remark that $V_j(t)$ acts nontrivially only on the two-dimensional one-body
subspace (one should also note $\text{tr}[V_j(t)]=0$).
In this subspace,
since $U_j$ is a $2\times2$ matrix, its logarithm can be expanded by the Pauli matrices, or the spin operators:
$U_j = e^{-i \bm{h}_\text{eff} \cdot \bm{S}_j T}$,
where $\bm{h}_\text{eff}$ is the expansion coefficients of the spin operators (we have ignored the constant term).
Note that this expression is valid on the total 4-dimensional Hilbert space.
Summing up all the terms, we obtain the Floquet Hamiltonian
\begin{align}
H_F^\text{el} = \frac{i}{T}\log U_F^\text{el} = H_0 + \bm{h}_\text{eff} \cdot \bm{S}. \label{sp:HF}
\end{align}
This result means that the stroboscopic time evolution under $H_\text{el}(t)$ is identical to 
that under $H_F$, namely the Hubbard model in the static magnetic field $\bm{h}_\text{eff}$.
The dynamical symmetry of the Floquet Hamiltonian~\eqref{sp:HF}, $[H_F, S^+_{\bm{h}_\text{eff}}] =  |\bm{h}_\text{eff}| S^+_{\bm{h}_\text{eff}}$, 
gives rise to the FDS:
\begin{align}
U_F^\text{el} S^+_{\bm{h}_\text{eff}} (U_F^\text{el})^\dag  = e^{-i\lambda T } S^+_{\bm{h}_\text{eff}},
\end{align}
where $S^+_{\bm{h}_\text{eff}}$ is a total spin raising operator along the effective magnetic field $\bm{h}_\text{eff}$,
and $\lambda = |\bm{h}_\text{eff}|$.
This FDS leads to the time crystals in the Hubbard model driven by the elliptically polarized ac field.
We note that, unlike the case of the circularly polarized ac field,  this time crystals cannot be understood by the unitary transformation to the rotating frame since the time-dependence remains even in the rotating frame.

Let us numerically demonstrate our results.
Figure~\ref{fig:elli} (a) shows the $\Delta$-dependence of $\lambda$ for $\omega = \pi$ and $B=4\pi/3$.
By tuning $\Delta$ as $\lambda=2\pi/3$ based on Fig.~\ref{fig:elli} (a),
we obtain the DTC dynamics with period $3T=6$, which is the least common multiple of $T=2\pi/\omega=2$ and $2\pi/\lambda=3$,
both in the dissipative and isolated systems (Figs.~\ref{fig:elli} (b) and (d)).
Moreover, the spin dynamics at the different sites are synchronized after the first relaxation due to dissipation and quantum thermalization.
On the other hand, for a not-fine-tuned $\Delta$, the synchronized DTQC dynamics appear both in the dissipative and isolated systems (Figs.~\ref{fig:elli} (c) and (e)).
These results are also confirmed by the Fourier analysis (Figs.~\ref{fig:elli} (f) and (g)),
where we see peaks at $\omega$ and $\lambda$
together with their harmonics and sum (difference) frequencies,
which implies the appearance of the time-crystalline nature.

\section*{S3. Time crystals in isolated integrable systems}

The key mechanisms of the time crystals in the isolated systems are the FDS and the equilibration to the stationary state with small time-fluctuation in the stroboscopic sense, although we have roughly used the term ``thermalization" in the main text.
Therefore, we expect that not only the nonintegrable systems exhibiting the thermalization but also the integrable systems not exhibiting the thermalization show the time-crystalline behavior in the thermodynamic limit due to the equilibration.

\begin{figure}[t]
\center
\includegraphics[width=0.9\columnwidth]{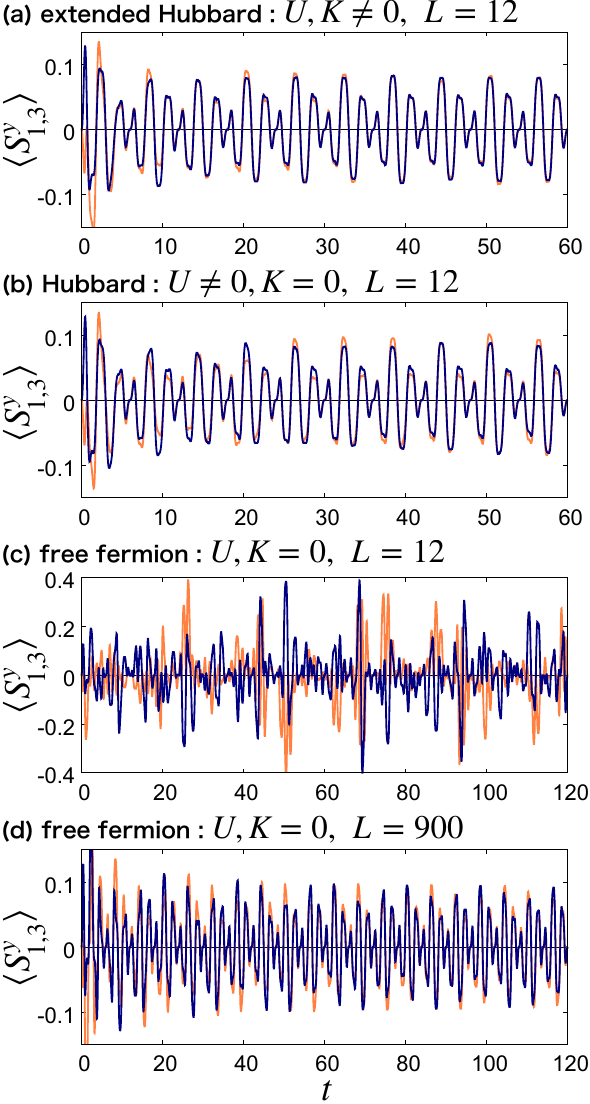}
\caption{
Time evolution of $\braket{S^y_{j}(t)}$ (navy: $j=1$, orange: $j=3$) in (a) extended Hubbard model,
(b) usual Hubbard model, and (c,d) free fermion model without dissipation.
The parameters are $J=1, \gamma=0, \omega=\pi, B=4\pi/3$, and $(U,K)=$ (a) $(3,1)$, (b) $(3,0)$, and (c,d) $(0,0)$. 
The system sizes are $L=12$ for (a,b,c) and $L=900$ for (d).
}
\label{fig:free}
\end{figure}

Figure~\ref{fig:free} shows the spin dynamics in (a) the extended Hubbard, (b) the usual Hubbard, and (c,d) the free fermion models driven by the circularly polarized ac field, respectively, for (a,b,c) $L=12$ and (d) $L=900$.
As shown in the figure,
the Hubbard model with $L=12$ (Fig.~\ref{fig:free} (b)) exhibits the time crystals as well as the extended Hubbard model (Fig.~\ref{fig:free} (a)) in spite of the integrability.
On the other hand, the dynamics of the free fermion system with $L=12$ (Fig.~\ref{fig:free} (c)) seems quite noisy.
This implies that the free fermion model does not equilibrate yet due to its high symmetry even for $L=12$.
As the system size increases, the equilibration becomes more accurate,
and the time-crystalline behavior finally appears for $L=900$ (Fig.~\ref{fig:free} (d)).

We note that the free fermion systems have an extensive number of (local) dynamical symmetries.
For example, the spinless fermions with the quadratic Hamiltonian $H=\sum_k \epsilon_k c_k^\dag c_k$
have the dynamical symmetries
$[H,c_k^\dag]=\epsilon_k c_k^\dag$, $[H,c_k^\dag c_l]=(\epsilon_k-\epsilon_l) c_k^\dag c_l$, and so on, where %$H=\sum_k \epsilon_k c_k^\dag c_k$ is 
$c_k$ ($c_k^\dag$) is an annihilation (creation) operator.
These dense and incommensurate frequencies $\epsilon_k$ can destroy the time crystals in general~\cite{Essler2016,Bokker2020}.
Nevertheless, as shown in Fig.~\ref{fig:free} (d), the multiple dynamical symmetries do not destroy the time-crystalline nature for the spins in our model.
However, they can be important in other situations such as the case with the particle number fluctuation,
where there can exist quantum coherence between $\ket{\psi}$ and $c_k^\dag\ket{\psi}$ protected by the dynamical symmetry $[H,c_k^\dag]=\epsilon_k c_k^\dag$.
The complete elucidation of the role of the multiple dynamical symmetries in free systems is an open question.

\section*{S4. GGE description of time crystals protected by Floquet dynamical symmetry}

Here we show the time-dependent GGE of the time crystals protected by the FDS 
in the isolated systems.

\subsection*{A. Preliminary: case of time-independent Hamiltonian}
For preliminary, let us consider the static Hamiltonian $H$ with the usual extended dynamical symmetry $[H,A]=\lambda A$ (this setup and its conclusion were discussed in Ref.~\cite{Medenjak2019}).
The extended dynamical symmetry leads to the stroboscopically-conserved quantity:$A_H(s+nT_\lambda)=A_H(s)$ in the Heisenberg picture for each $s\in[0,T_\lambda)$ with $T_\lambda=2\pi/\lambda$ and $n\in\mathbb{N}$.
This unconventional conserved quantity prevents the system from relaxing to the stationary state and gives rise to the time-dependent GGE with period $T_\lambda$: 
\begin{align}
\rho(t) = \frac{\exp[-\sum_i \beta_i(t) Q_i - \mu(t)A - \mu^\ast(t)A^\dag]}{Z_t}.
\end{align}
Here $Z_t$ is the periodic partition function,
$\mu(t)$ is the periodic chemical potential for $A$,
and $Q_i$ and $\beta_i(t)$ are the conventional local conserved quantities of $H$
and their periodic generalized inverse temperatures (the time-dependence of $\beta_i(t)$ could stems from the noncommutativity between $Q_i$'s and $A$, but in fact, $\beta_i(t)$ does not depend on time as shown below).
The chemical potential $\mu(t)$ and the generalized inverse temperatures $\beta_i(t)$ are determined by 
the following relations for any $t$:
\begin{align}
\begin{aligned}
\braket{\psi(t) | Q_i | \psi(t)} &= \text{tr}[ Q_i \rho(t)], \\
\braket{\psi(t) | A | \psi(t)} &= \text{tr}[ A \rho(t)].
\end{aligned}
\label{sp:cond}
\end{align}
These relations are satisfied by $\beta_i(t) = \beta_i$ and $\mu(t) = \mu e^{-i\lambda t}$, where $\beta_i$ and $\mu$ are time-independent quantities satisfying Eqs.~\eqref{sp:cond} at $t=0$.
To prove this, we have used $U^\dag(t) A U(t) = e^{i\lambda t} A$, $U^\dag(t) Q_i U(t) = Q_i$,
and $\rho(t) = U(t) \rho(0) U^\dag(t)$ if $\beta_i(t) = \beta_i$ and $\mu(t) = \mu e^{-i\lambda t}$.
Then, $Z_t $ is also  time-independent because of $\rho(t) = U(t) \rho(0) U^\dag(t)$.
As a result, we obtain 
\begin{align}
\rho(t) = \frac{\exp[-\sum_i \beta_i Q_i - \mu(t)A - \mu^\ast(t)A^\dag]}{Z}, \label{sp:normGGE}
\end{align}
where $\mu(t)=\mu(0)e^{-i\lambda t}$.

\subsection*{B. Time-dependent GGE for time crystals protected by FDS}

Now we consider the time-periodic Hamiltonian $H(t)=H(t+T)$ with the FDS.
To obtain the GGE,
we invoke a theoretical trick of
introducing a virtual time evolution under
the Floquet Hamiltonian $H_F=(i/T)\log U_F$.
Namely, the evolution from time $0$ to $t$ is given by $e^{-iH_F t}$ 
in this virtual evolution.
While the real evolution generated by $H(t)$ is different from the virtual one
at most times,
they coincide with each other at the stroboscopic times $t=nT$ ($n\in\mathbb{N}$).
From now on, we call the density matrix obtained by the evolution under $H(t)$ as $\rho_\text{TC}(t)$ and that under $H_F$ as $\rho_F(t)$.
Given that they share the initial condition $\rho_\text{TC}(0)=\rho_F(0)$,
we have
\begin{align}
\rho_\text{TC}(nT) = \rho_F(nT).\label{sp:real_virtual}
\end{align}

Here we assume that
the Floquet Hamiltonian $H_F$ has the same local conserved quantities as $U_F$ and the extended dynamical symmetry, 
 \begin{align}
[H_F,Q_i]=0, \,\,\, [H_F,A]=\lambda A, \label{sp:assumption}
\end{align}
where $Q_i$ are local (or few-body) conserved quantities of $U_F$: $[U_F,Q_i]=0$.
Note that the assumption~\eqref{sp:assumption}
implies the FDS and holds in the Hubbard models under the circularly and elliptically polarized ac fields
discussed in the main text and Sec.~S2.
Since Eq.~\eqref{sp:assumption} means the extended dynamical symmetry,
the argument in the previous subsection applies to the present virtual evolution,
and we have 
$\rho_{F}(t) = \exp[-\sum_i \beta_i Q_i - \mu(t)A - \mu^\ast(t)A^\dag]/Z$
(see Eq.~\eqref{sp:normGGE}).
Here $Z$ is the partition function,
$\mu(t)=\mu e^{-i\lambda t}$ is the periodic chemical potential for $A$,
$Q_i$ are the local conserved quantities of $H_F$ (and, hence, $U_F$),
and $\beta_i$ are the generalized inverse temperatures for them.

The GGE thus obtained for the virtual evolution
gives that for the real evolution of interest
at the stroboscopic times through Eq.~\eqref{sp:real_virtual}.
To obtain the GGE at different times,
we act the time evolution operator $U(t_0)$ from $nT$ to $t_0+nT$,
obtaining the time-dependent GGE $\rho_\text{TC}(t_0+nT)=U(t_0)\rho_\text{TC}(nT)U^\dag(t_0)$:
 \begin{align}
\rho_\text{TC}(t) 
= \frac{\exp \left[ -\sum_i \beta_i Q_i(t) - \mu(t) \tilde{A}(t) - \mu(t)^\ast \tilde{A}^\dag(t) \right]}{Z}. \label{sp:tFGGE}
\end{align}
Here, $Q_i(t)=Q_i(t_0)=U(t) Q_i U^\dag(t)$ is periodic conserved quantities of $H(t)$,
$\tilde{A}(t)\equiv e^{i\lambda t}U(t) A U^\dag(t)$
is a periodic FDS operator,
and $\mu(t)=\mu e^{-i\lambda t}$ is the periodic chemical potential for $A$.
This result~\eqref{sp:tFGGE} is consistent with Eq.~(9) in the main text, which is derived by considering the stroboscopically-conserved quantities of the DTC.

\begin{figure}[t]
\center
\includegraphics[width=0.9\columnwidth]{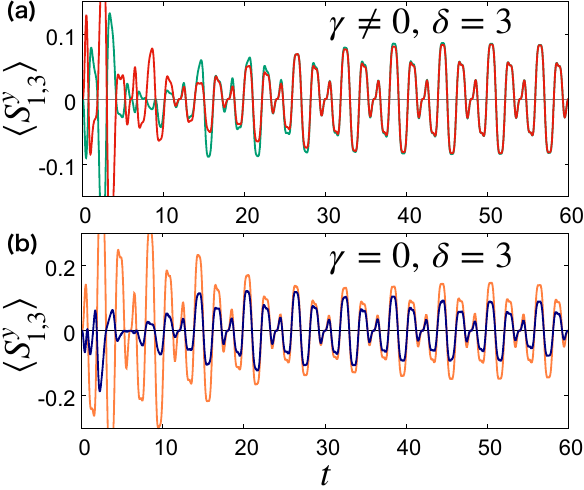}
\caption{
(a) Time evolution of $\braket{S^y_{j}(t)}$ (red: $j=1$, green: $j=3$) in dissipative Hubbard model with disordered potential.
The parameters are $L=6, J=U=K=1, \gamma=0.1, \delta=3, \omega=\pi$, and $B=4\pi/3$.
(b) Time evolution of $\braket{S^y_{j}(t)}$ (orange: $j=1$, navy: $j=3$) in isolated Hubbard model with disordered potential.
The parameters are $L=12, J=U=K=1, \gamma=0, \delta=3, \omega=\pi$, and $B=4\pi/3$.
Both in (a) and (b), we have used one realization for the disorder potentials $\{\epsilon_j\}$ rather than the ensemble average,
and the realizations for (a) and (b) are different.
}
\label{fig:disorder}
\end{figure}

In deriving Eq.~\eqref{sp:tFGGE} for the DTC and DTQC, we have assumed the conditions~\eqref{sp:assumption}.
On the other hand, 
the derivation of Eq.~(9) in the main text does not need any assumptions although it is limited to the DTC case.
It is an open problem
whether the assumption~\eqref{sp:assumption} really needs or not in the DTQC case.

\section*{S5. Robustness against disorder}

Here, we show that
the time crystals protected by the FDS are robust against perturbations
as long as the FDS is preserved, although the synchronization in the isolated systems may become imperfect.
To illustrate this, 
we consider the disordered onsite potential
\begin{align}
V_d = \sum_j \epsilon_j n_j,
\end{align}
where $\epsilon_j$'s are independent random variables following the uniform distribution over $(-\delta,\delta)$.
This disorder respects the FDS due to the spin-SU(2) symmetry, 
but breaks the spatial translation symmetry.

In the presence of the dissipation $L_j=n_j$, not only the time-crystalline nature but also the synchronization are robust against the disorder.
Figure~\ref{fig:disorder} (a) shows the dynamics of $\braket{S_j^y}$ ($j=1,3$) in the disordered Hubbard model with dissipation for the DTC phase, $\omega = \pi$ and $\lambda=2\pi/3$, and thus the period is $3T$.
As in the case without disorder (Fig.~2 (b) in the main text), the two spin dynamics become synchronized to oscillate with period $3T$.
This synchronization is caused by the dissipation $L_j=n_j$ suppressing the particle number fluctuations to realize the translationally symmetric state.

On the other hand, in the absence of dissipation,  
the perfect synchronization does not occur,
but the time-crystalline nature persists.
Figure~\ref{fig:disorder} (b) shows the dynamics of $\braket{S_j^y}$ ($j=1,3$) in the disordered Hubbard model without dissipation for the DTC phase with period $3T$.
Unlike the dissipative system, the amplitudes of the two spin dynamics are different, meaning that the synchronization is imperfect.
However, the rhythms of the oscillations are synchronized, and the time crystal with period $3T$ appears.
This spatial inhomogeneity can be understood by quantum thermalization.
Since a state thermalizes to an equilibrium state of the disordered Hamiltonian, the realized state is not translationally symmetric.

Finally, we note that many-body-localized (MBL) systems have an extensive number of local dynamical symmetries in common with the free systems (see Sec.~S3)~\cite{Khemani2019},
which can destroy the time crystals due to the dense and incommensurate frequencies.
However, in the Hubbard model, it is known that the disordered onsite potential never leads to the MBL in the spin sector since the onsite potential acts on the up and down spins equally, although it does in the charge sector~\cite{Prelovsek2016}.
Thus, as for the local spin dynamics, the time-crystalline nature persists even in the presence of the strong disorder as shown in our results. 
The complete elucidation of the role of the local dynamical symmetries in the MBL systems 
is an open question and requires more thorough analysis.

\end{document}